**Title:** Predicting the ecological outcomes of global consumption

**Authors:** Payal Bal[1], Simon Kapitza[1], Natasha Cadenhead[2], Tom Kompas[1,3], Pham Van Ha[4], Brendan Wintle[1]

**Affiliations:**

1. School of BioSciences, The University of Melbourne, Melbourne, VIC 3010, Australia
2. University of Queensland.
3. Centre of Excellence for Biosecurity Risk Analysis, University of Melbourne, Melbourne, VIC 3010, Australia
4. Crawford School of Public Policy, Crawford Building (132), Lennox Crossing, Australian National University, ACT 2601, Australia

**Abstract:** Mapping pathways to achieving the sustainable development goals requires understanding and predicting how social, economic and political factors impact biodiversity. Trends in demography, economic growth, regional alliances and consumption behaviours can have profound effects on the environment by driving resource use and production. While these distant socio-economic drivers impact species and ecosystems at global scales, for example by driving greenhouse gas emissions and climate change, the most prevalent human impacts on biodiversity manifest through habitat loss and land use change decisions at finer scales. We provide the first integrated ecological-economic analysis pathway capable of supporting both national policy design challenges and global scale assessment of biodiversity risks posed by socio-economic drivers such as population growth, consumption and trade. To achieve this, we provide state-of-the-art integration of economic, land use, and biodiversity modelling, and illustrate its application using two case studies. We evaluate the national-level implications of change in trading conditions under a multi-lateral free trade agreement for the bird biodiversity of Vietnam. We review the implications for land-use and biodiversity under coupled socio-economic (Shared Socioeconomic Pathways) and climate (Resource Concentration Pathways) scenarios for Australia. Our study provides a roadmap for setting up high dimensional integrated analyses foe evaluating global priorities for protecting nature and livelihoods in vulnerable areas with the greatest conflicts for economic, social and environmental opportunities.

Population growth, consumption and trade are direct socio-economic drivers of land-use change, climate change and invasive species, which in turn determine where species and ecosystems persist (IPBES 2016). The UN Sustainable Development Goals (SDGs)[1] explicitly acknowledge people's reliance on nature and our role in determining outcomes for nature, as decades of scientific evidence link increased consumption and commensurate land use change with biodiversity loss and the degradation of nature (Foley *et al.* 2005; Newbold *et al.* 2015). The imperative to better account for the adverse consequences of socio-economic drivers now also permeates key regional and national strategies for nature (The State of Victoria Department of Environment, Land, Water and Planning 2017; Commonwealth of Australia 2019) as well as global policy frameworks and targets for conservation (Steffen *et al.* 2015). While the links between socio-economic drivers and environmental degradation can be obvious in hindsight, their prediction is difficult. The absence of credible predictive modelling represents a critical bottleneck in the development of sustainable macro-policy. Mapping plausible and effective policy pathways toward a more sustainable relationship between national wealth and nature, the so-called 'green economy' (Sukhdev *et al.* 2015), at national and global scales is extremely complex (Bryan *et al.* 2015; Bai *et al.* 2016). Policies that appeal to a green economy may be controversial because of the unpalatable trade-offs they can invoke between established economic interests and uncertain consequences for nature. This study provides the basis for credible quantitative characterisation of the outcomes for nature under plausible macro-scale socio-economic scenarios.

Distant social and economic forces impact species and ecosystems at global scales, for example through the impact of resource production on greenhouse gas emissions and climate change, make for a truly tele-coupled system (Liu *et al.* 2013; Torres *et al.* 2017). The most prevalent human impacts on biodiversity, however, manifest through habitat loss and degradation arising from land use decisions at finer scales (Foley *et al.* 2005; Maxwell *et al.* 2016). Economic trade facilitates the flow of goods and services between countries facilitating the emergence of global human consumption and production patterns and is a key driver for shifting national land use profiles (Weinzettel *et al.* 2013). For instance, trade enables technology exchange to improve efficiency of production thereby reducing the pressure to convert natural habitats to agricultural or industrial land. Conversely, new trading agreements can drive the demand for goods and services that displace human activities, e.g. agriculture and urban settlements, into areas that may be important for species persistence (Visconti Piero *et al.* 2015). The last 20 years alone, have seen a tenfold increase in total food exports across the world (United Nations Statistics Division 2017). Although the demand for key commodities, and the places they are most likely to be derived is somewhat predictable (e.g. palm oil is most likely to be grown on tropical land with low economic opportunity cost, such as cleared rainforest), the inherent unpredictability of emerging geopolitical forces that influence trading alliances, or emergent technologies that generate demand for new commodities can lead to dramatic and unforeseen changes in global land use patterns. Such changes can have a profound influence on biodiversity conservation outcomes.

---

[1] www.un.org/sustainabledevelopment/biodiversity/

Until now, studies describing the links between consumption, trade and biodiversity have failed to provide spatially explicit assessments of the impacts of socio-economic drivers on biodiversity persistence below the national level (e.g., Lenzen *et al.* 2012; Moran & Kanemoto 2017). They have relied on reasonable but coarse first-order approximations of impacts on biodiversity by considering the spatial extent of the drivers and their overlap with the location of habitats for vulnerable biodiversity (Lenzen *et al.* 2012; Newbold *et al.* 2015). Often studies have used existing land-use projections to evaluate the effect of land-use change on biodiversity under global climate or socio-economic scenarios (Jetz, Wilcove & Dobson 2007; Pereira *et al.* 2010; Newbold 2018), making it hard to evaluate bespoke socio-economic scenarios or policies (but see Visconti *et al.* (2016) for an evaluation under assumptions for food consumption, annual crop yields and income disparity using the IMAGE model). Lack of spatially explicit predictions about where conflict between nature and land use will play out next impairs global efforts to protect biodiversity under future economic growth scenarios.

Exponential growth in data and information driven by public data repositories and new sensing technologies provide greater insights into the workings of the economy (Ha *et al.* 2017), the drivers of land use (Bryan *et al.* 2015), and the dynamics of environmental and biodiversity change. These new data sources combined with statistical modelling and synthesis methods provide the foundations for dramatically improved understanding about the dynamics and behaviour of socio-ecological systems on which prediction can be based. While it's difficult to anticipate unpredictable events, scenario analysis can be used to develop robust policy under plausible global futures. For instance, the shared socio-economic pathways (SSPs) are a widely used set of scenarios within which more specific analyses of the impacts of policies and economic shocks on land use and biodiversity can be analysed (O'Neill *et al.* 2017). We provide the first integrated analysis pathway capable of supporting both national policy design challenges and global scale assessment of biodiversity risks posed by future change in socio-economic drivers such as population growth, consumption and trade.

To achieve this ambitious aim, we provide state-of-the-art integration of economic, land use, and biodiversity modelling, to track the pathways from global consumption to production, and to outcomes for species and ecosystems in an otherwise complex, non-linear and stochastic system. High resolution data describing global trade in commodities is used in dynamic economic models and coupled with spatially-explicit land use, climate change, and species-level biodiversity models to assess the impacts of climate, consumption and socio-economic trends (e.g., demography or economic shocks like new trade agreements) on nature (Kapitza *et al.* 2020). This ensures that the spatial, temporal and thematic changes represented in future climate and socio-economic scenarios match the operational scale at which biodiversity respond to these changes (Titeux *et al.* 2016). Our goal is to provide new insights into economic versus ecological opportunities and trade-offs when making decisions at national, regional and global scales. This will enable governments and large conservation organisations prioritise investment for the protection of places and livelihoods that will need it the most under current trajectories of environmental and economic change.

### *An integrated modelling framework*
Our approach down-scales global economic and demographic patterns to local (<1ha) biodiversity impacts via a land-use change model, to allow predictions at annual, decadal and millennial timescales. Building on recent advances in economic modelling (Ha & Kompas

2016), land-use trade-off modelling (Bryan *et al.* 2016) and ecological modelling (Wintle *et al.* 2011; Morán-Ordóñez *et al.* 2017), our approach provides unprecedented, high resolution, national and global scale spatial analyses of the tele-coupling of socio-economic and environmental systems. With the aid of cutting-edge, high-performance computing infrastructure, we are on the cusp of an analytical pipeline capable of applying integrated ecological-economic models to a suite of global- and national-scale policy modelling challenges. Below we describe the components of our framework in detail.

*Scenarios, policy options and economic shocks:*

Scenarios articulate a broad set of operating conditions (or assumptions) that set the context for more specific modelling and consideration of policy options (IPBES 2016). These may be exploratory in nature that provide alternative trajectories for environmental and socio-economic factors that currently impact biodiversity or are expected to do so in the future (Liu *et al.* 2013; Rosa *et al.* 2020) . For instance, climate scenarios represented by the resource concentration pathways (RCPs), describe pathways of greenhouse gas emissions leading to specific changes in surface temperature. Complementing these are the shared socio-economic pathways (SSPs) that provide five distinct narratives about the future of the world, exploring a wide range of plausible trajectories of population growth, economic growth, technological development, trade development and implementation of environmental policies (O'Neill *et al.* 2017). These provide alternative contexts for mapping future land use change and biodiversity conflict. While O'Neil avoids placing SSPs on a scale, the descriptions progress from high focus on sustainability, equity and low emissions (SSP1), through to unmitigated industrialisation and emissions growth (SSP5). Each SSP aligns broadly with one or two global carbon emission scenarios known as representative concentration pathways (RCPs), allowing climate change scenarios to be built into SSPs (Riahi *et al.* 2017).

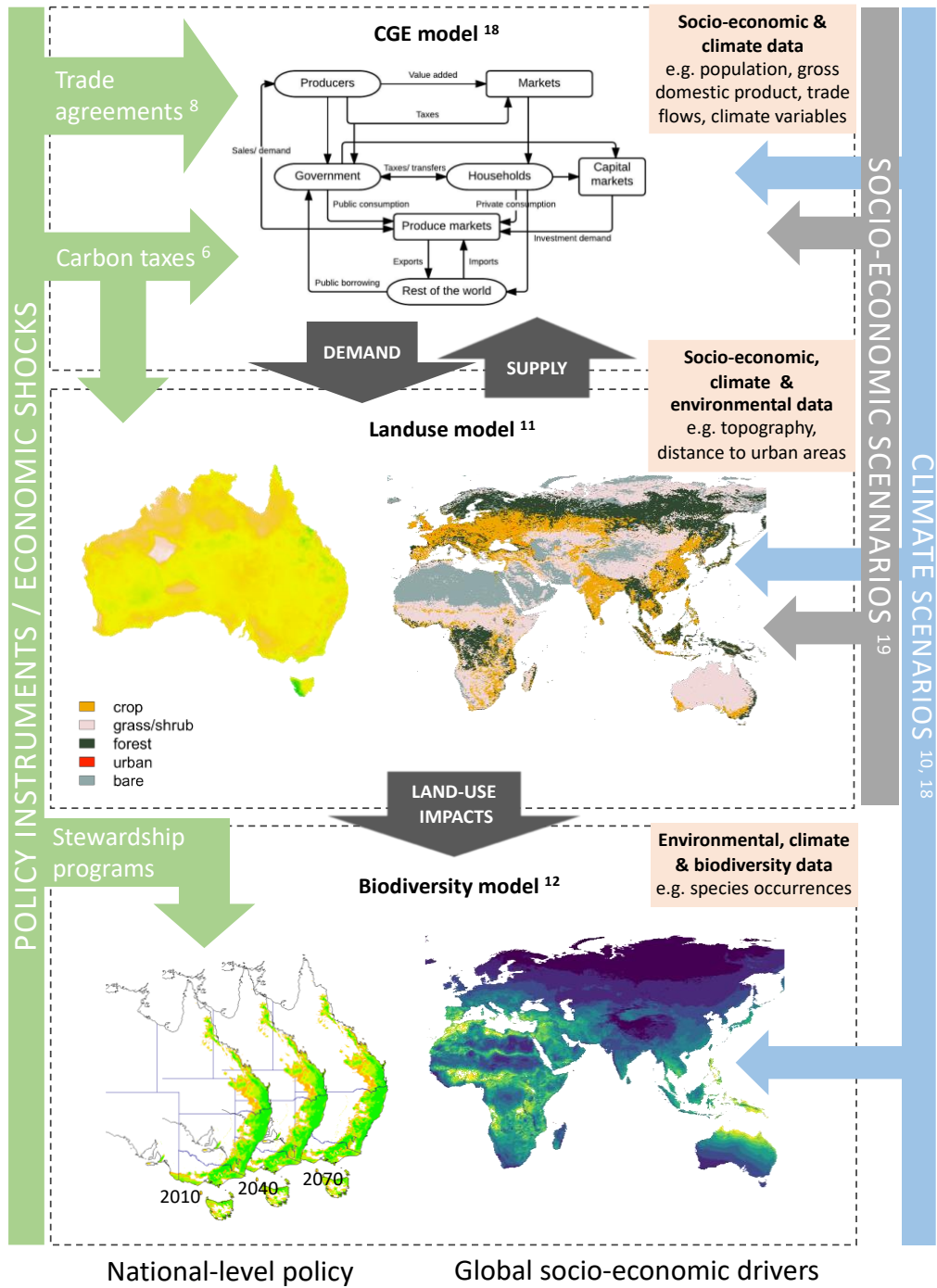

Figure 1. An integrated ecological-economic modelling framework to assess the impacts of global socio-economic drivers on species and ecosystems. Scenarios, policy options and economic shocks are parameterised at the national or global level using publicly available data on the flow of goods and services (Kompas & Van Ha 2019), land-use (Hurtt *et al.* 2011), and human demographic and environmental variables (Pereira *et al.* 2013). Trade agreements, carbon taxes or land stewardship programs are examples of economic shocks or policy instruments that can be addressed by the framework. Shared socio-economic pathways (SSPs) are a widely used set of scenarios describing futures in socio-economic drivers (van Vuuren *et al.* 2011; O'Neill *et al.* 2014, 2017). Representative

concentration pathways (RCPs) are scenarios that link human activities to greenhouse gas and aerosol emissions (Riahi *et al.* 2017), and can directly and indirectly influence economy, land use and biodiversity models. Logical coupling of SSPs and RCPs is necessary for integrated ecological-economic predictions. Computable global equilibrium (CGE) models translate specific scenarios and policies into sectoral demands (e.g. paddy, oil seeds, forestry) over the long term (Kompas, Pham & Che 2018). Projected demand from economic sectors are linked to land-use change (<1ha) through the land-use model10. Spatial distributions of 1000s of species are modelled as a function of static topographic variables, and dynamic land use and climate variables (Kapitza *et al.* 2020). Biodiversity response is summarized using indices derived from stacked species distributions.

Alternately, target-seeking scenarios may be used, for instance to explore pathways to achieve the 2050 strategic vision of the Convention on Biological Diversity (Kok *et al.* 2014) or the Sustainable Development Goals set out in the United Nations 2030 agenda (Messenger 2017). Qualitative and quantitative approaches, including integrated assessment models, systems models, and expert knowledge, can be used to identify impacts the expected impacts of specific policy alternatives for reaching endpoints in case of target-seeking scenarios or make predictions for biodiversity under the exploratory scenarios (Nicholson *et al.* 2019).

Trade agreements, carbon taxes or land stewardship programs are examples of economic shocks or policy instruments that can be addressed by the framework. Trade agreements can be designed by simulating the long-run effects of bilateral or multilateral trade agreements resulting in changes in commodity flows between participating countries. Alternative trade scenarios can be incorporated by varying tariffs or import taxes, non-tariff barriers such as regulations and domestic subsidies, and foreign direct investments within economic models (Fuss *et al.* 2015; Corong *et al.* 2017; e.g., Li, Scollay & Gilbert 2017). Here, the counterfactual or business-as-usual scenario – i.e., when the policy is not implemented – must also be stated. Scenarios outside those which are currently realistic – extreme cases, or best-or-worst case scenarios – may also be explored. There are also inherent assumptions about the openness of trade or trade barriers within each of the SSPs that can be further explored. For instance, SSP 3 and SSP 4 represent scenarios with lowest levels of international co-operation and trade (O'Neill *et al.* 2017).

> **Box 1: Free trade agreements**
>
> Very little is known about how social, political or economic changes driven by trade agreements amongst countries will impact the environment (Pace & Gephart 2017). For instance, free trade agreements (FTAs) reduce or remove trade and investment barriers between countries, providing substantial income gains worldwide beyond those incurred by participating countries alone, and alter the global supply and production chains. This not only transforms the way commodities are produced, exchanged and consumed, but also changes the location and scale of both social and environmental impacts (Wiedmann & Lenzen 2018). FTAs such as the Trans-Pacific Partnership (TPP), an agreement among twelve Pacific Rim countries on trade and economic policy would made up nearly 1% of the global population and approximately 40% of global gross domestic product had it been ratified by the United States (Countryman, Warziniack & Grey 2018). New tariff schedules under currently debated 'protectionist' trade agreements following Brexit or the revised North American Free Trade Agreement (NAFTA) are expected to lead to significant changes in trade flows and consumer behaviour, leading to the emergence of new trade partnerships and putting greater focus on domestic markets. As a result, these decisions have the potential to drive new areas of expansion across different sectors of the economy, which can translate to large changes in global land use along with potentially large environmental impacts (Liu, Newell & White 2015; Countryman, Warziniack & Grey 2018).
>
> FTAs can be economically beneficial to a country, however they have also been criticised based on concerns about weakening local production economy, job loss, labour and human rights violations, and poor environmental outcomes. Disapproval of FTAs from environmental organisations is usually focused around investor-state dispute settlement, provisions that allow foreign investment companies to sue governments for legislating for environmental protections that might harm company profits. While some economists feel that these dangers are overblown (Schott 2016), this provision is likely to remain an ongoing concern. Even without these legal impediments, very little is known about how the economic changes effected by FTAs will impact the environment in their own right, and this question has largely been absent from debate.

*Economic model:*

The first sub-model within our framework is a computable general equilibrium (CGE) which is a global economic model used to evaluate the impacts of large-scale, multilateral policy or other form of economic shocks, such as a new tax or trade agreement (Ha *et al.* 2017; Kompas & Van Ha 2019), on different sectors of the global or regional economies. CGEs can predict patterns and changes in trade, production, and consumption of commodities in response to economic shocks, such as political treaties, economic taxation schemes, or trade agreements (Piermartini & Teh 2005; Ha *et al.* 2015) . These models represent the economy as a complete system of interdependent components, including industries, households, investors, governments, importers and exporters, such that perturbation in any one component can have flow-on effects throughout the system (Dixon *et al.* 1992). Mathematically, they are formulated as a set of equations which describe the economic behaviour of a representative agent from each component of the model economy, e.g. a typical household, producer, or an importer or exporter under explicit institutional and technological constraints. Behavioural equations are represented as explicit constrained-optimization problems (Dixon *et al.* 1992). CGEs rely on two types of data: input-output tables and behavioural parameters. Input-output tables record the commodity flows between components of the economy. Behavioural parameters describe how agents respond to changes in the system, e.g. change in consumption of a commodity by a household in response to change in incomes and consumer prices (Dixon *et al.* 1992).

The Global Trade Analysis Project (GTAP) is a multiregional, multi-sector CGE model, containing information on trading partnerships. The underlying GTAP database is the largest available input-output and trade database, currently including 140 countries and 57 commodities (Aguiar, Narayanan & McDougall 2016). The model uses two types of equations: (1) accounting relationships which ensure that receipts and expenditures of every agent in the economy are balanced, and (2) behavioural equations based upon microeconomic theory (Brockmeier 2001).

*Land-use model:*

Land-use and land-use change are key drivers of species persistence and ecosystem condition (Foley *et al.* 2005). Coupling CGE projections to spatially-explicit land-use information is an essential step in our framework as it downscales global trade flows to local level decisions for land-use and subsequently to biodiversity impacts. Approaches to simulate near future land-use patterns in response to economic factors (i.e., commodity demands) encompass modelling the projected demand for individual land-use types or classes, as well as the spatial interactions and competition between land use classes to determine the allocation of change in land-use classes across the study area (Briassoulis 2000; Verburg *et al.* 2004). When demand for land-use types is largely determined by forces that are exogenous to the land allocation, 'top-down' approaches to determine the total area of the individual land-use type is followed by spatial allocation of the change in land-use type to individual grid cells (Verburg & Overmars 2009b). This is the case for agricultural land where the dynamics of change in area required are governed largely by changes in regional or global demand and market conditions of commodities like food, feed and energy. Conversely, for natural habitats and forests, aggregate demand cannot always be determined except when driven by wood demand or policy designation. In these cases, a 'bottom-up' approach is used where aggregate change in area is determined by the dynamics of the other demand-driven land-use types. Subsequently, spatial allocation is determined by allowing transitions according to conversion rates or elasticities specific to individual land-use classes (Verburg & Overmars 2009b). We make use of these principles for our analysis.

A rich set of national-scale land use allocation models now exist that can predict optimal or likely land uses in future times based on a set of commodity demand and household behavioural assumptions (Bryan *et al.* 2015). The Land Use Trade-Off model (LUTO; Bryan *et al.* 2015) has been used to evaluate national-scale potential for land use change toward carbon sequestration. It is an example of a highly parameterised model that utilises data about land-holder behaviour to help inform land use elasticities and how they vary with social and demographic parameters. This approach can be loosely described as a mechanistic model of land use choices, and contrasted with more pattern-based, statistical modelling approach for determining land-suitability and allocation under demand scenarios (Verburg & Overmars 2009b; Kapitza *et al.* 2020). Pattern-based approaches sacrifice some detail and predictive specificity in order to be tractable over larger areas and to support global analyses (Kapitza *et al.* In prep.).

*Biodiversity model:*

Species distribution modelling is one of the most rapidly developing and highly cited topics in environmental science (Sequeira *et al.* 2018). The third sub-model in our framework links the spatial dynamics of land-use change with species distributions and habitat suitability. Correlative pattern analysis, usually based on species distribution, species richness and compositional turnover models (Ferrier *et al.* 2007; Lawler *et al.* 2009) remain the state-of-the-art in biodiversity change analysis at all scales above local (e.g. provincial, national, global).

We use correlative species distribution models (Phillips, Anderson & Schapire 2006; Renner *et al.* 2015) to delineate the habitat range and characterise spatial variation in the suitability of the habitat for individual species based static topographical, and dynamic land use and climate variables (Stanton *et al.* 2012). This enables us to predict the response of individual species to socio-economic drivers of change that influence these two types of independent variables. These models are species specific and allow an assessment of response of individual species to drivers of change or of aggregated response of biodiversity as a whole through the use of indices such as mean species abundance and richness (Ferrier *et al.* 2007; Hui *et al.* 2013). Estimates of an aggregate index of biodiversity composition through the use of Generalised Dissimilarity Modelling (GDM), an approach to examine patterns of diversity (Ferrier *et al.* 2002, 2007), may also be used but these do not provide species-level predictions required to inform national policies on species management or red-list category designation.

Recent studies evaluating spatial biodiversity patterns under climate and/or land-use change have used extent of occurrence analysis (Jetz, Wilcove & Dobson 2007), SDMs (Morán-Ordóñez *et al.* 2017; Newbold 2018), or bespoke models to estimate the extent of suitable habitat (Visconti Piero *et al.* 2015). Integrated assessment models (IAMs) such as PREDICTS (Hudson *et al.* 2017) and GLOBIOM (Havlík *et al.* 2014) are also being increasingly used in global biodiversity assessments.

*Model coupling:*

Co-dependencies, and non-linear interactions between economy, land-use and ecology determine the realised outcomes in all three sectors. Existing partial equilibrium analyses (without dynamic feedbacks between economy and land use), targeting the relationship between carbon price and land-use, assure us that model integration is possible (Bryan *et al.* 2015). Our framework links an economic, land-use change and biodiversity model to downscale global consumption and production dynamics to the level of local land use and its impacts on biodiversity. Loose coupling allows the output of one model to become the input of the successive model, but feedbacks are not yet implemented. Economic predictions from a CGE provide information on consumption patterns through aggregate, national-level demand for traded commodities under alternate scenarios. A land-use model predicts how the current configuration of land-use will change through time to meet the requirement of total area of different land-use classes needed for the production of individual commodities as set by the economic predictions. The time-series of land-use maps thus generated become one of the variables in the biodiversity models in addition to other environmental and anthropogenic variables such as climate, topography and disturbance. The biodiversity model can then predict where the landscape is most suitable for the occurrence or persistence of species of interest, for instance by using species distribution models or dynamic landscape metapopulation model.

*Model uncertainty & validation:*

Each modelling component comprises its own set of parameters and assumptions, which introduce uncertainty in model outputs. For instance, climate predictions vary greatly in the range of projected temperature increase over the next 100 years under the different RCP scenarios (IPCC 2014) and the predictive performance for species distribution models varies with observed species ranges and to the environmental coverage of occupancy data (Moran-Ordonez et al. 2017). Although separate uncertainty analyses provide some insight into the influence of various model components on the predictions (Woolley *et al.* 2017; Thuiller *et al.* 2019), a global sensitivity analysis comparing assumptions in an integrated framework is needed to identify the most problematic and influential sources of uncertainty (Conlisk *et al.* 2013). Combining predictive outputs from multiple instances of the a method or from multiple methods or algorithms, known as 'ensemble' models do not consistently prove to have superior predictive performance to individual models (Hao *et al.* 2019).

Models can be tested and evaluated using holdout validation to historical data. In the case of the CGE & land-use model coupling, historical patterns of land-use change can be used to test how well the model can predict the observed changes based on what was known about the economic and trade settings of the day. Similarly, for the land-use & biodiversity model coupling, the biodiversity model can be tested by fitting models to historical land-use data in a sample of landscapes with known land-use change and species persistence histories. A range of extinct, stable, and in-decline species can be used for model testing, to see how well the models can predict species' statuses at the current time.

**Data and methods**

We illustrate the proposed framework by applying it to two case studies. In the first case study, we evaluate the long run effects of an economic shock, i.e. a free trade agreement, for the bird biodiversity of Vietnam under an extreme climate change scenario (RCP 8.5). For the second case study, we predict the biodiversity response of Australian birds under coupled SSP-RCP scenarios to represent plausible global socio-economic and environmental futures for the world. Considerable resources for data selection and harmonisation were needed to achieve the desired spatial resolution and detail for representing the desired scenarios for each case study and for evaluating the past, present and future impacts on biodiversity therein (Hurtt *et al.* 2011; Fuchs *et al.* 2013).

*Scenario development:*

For the first case study, we explored the national-level implications of the multi-lateral trade liberalisation scenario akin to the revised Trans-Pacific Partnership (TPP) free trade agreement, an agreement originally among twelve Pacific Rim countries on trade and economic policy, for the bird biodiversity of Vietnam. Based on a recent analysis of the long-run effects of the TPP for Vietnam (Ha *et al.* 2017), we defined two plausible trade scenarios: a *TPP scenario* under which tariffs are sequentially removed or reduced between participating countries through time as per the TPP agreement and, and a *business-as-usual* scenario in which tariffs remain unchanged as per existing regional FTAs excluding the TPP

(see Ha *et al.* 2017 for details),. Scenarios were run from 2017 onwards under the most pessimistic RCP 8.5 emissions scenario to account for climate change (WorldClim Version 1.4).

For the Australian case study, we explore three scenarios combinations: (1) low change SSP 1 – RCP 4.5, moderate change SSP 2 – RCP 6, and (3) high change SSP 3 – RCP 8.5, to represent a continuum from low to high challenges for adaptation and mitigation and corresponding gradients in change in key scenario elements (GDP, population growth and technological development) (Figure x).

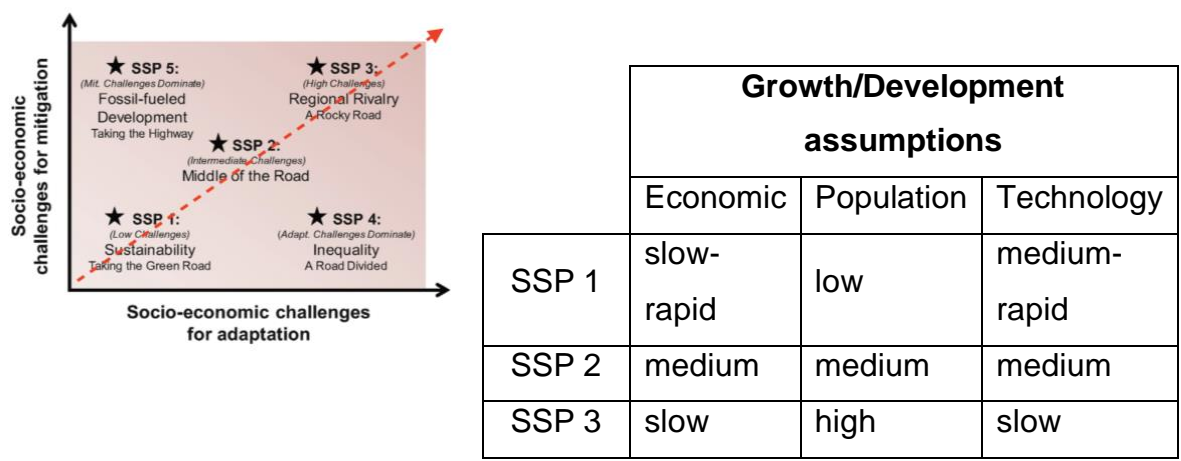

| | Growth/Development assumptions | | |
|---|---|---|---|
| | Economic | Population | Technology |
| SSP 1 | slow-rapid | low | medium-rapid |
| SSP 2 | medium | medium | medium |
| SSP 3 | slow | high | slow |

Figure x: Five shared socio-economic pathways (SSPs) from O'Neill *et al.* (2017)and their corresponding assumptions for key scenarios elements (van Vuuren & Carter 2014).

The baseline for SSPs was based on economic and population growth predicted for SSPs (IISAA's prediction) over 2011-2100. This generates higher temperature then those predicted for the RCPs considered within our scenario definitions because GHG limitations are not considered yet. To account for this and predict temperature rise concomitant with each SSP-RCP pairing, we make assumptions on fuel efficiency such that SSP 2 corresponds to the current rate of approximately 1.78% annually. The rate for other SSP are adjusted relative to SSP, i.e. higher for SSP1 and lower for SSP3, respectively.

*Economic model:*

Using the GTAP model (version 6.2) and GTAP database version 9, Ha and colleagues (2017) predicted the total amount of change in commodity demands in 2070 for 35 commodities/sectors (e.g. paddy, oil seeds, forestry) under the trade conditions of the BAU and TPP for 44 regions for Vietnam and under the coupled SSP-RCP scenarios for Australia. Further details on data and model used can be found in Ha *et al.* (2017). For each case study, we considered a subset of ten commodities that could be intuitively linked to five land-use classes (Table 1).

(a)

| Land-use classes | Description |
|---|---|
| Cropland | Land used for crop production (e.g., wheat, rice, oil seeds) |
| Grassland & shrubland | Native as well as non-native grasslands, including open, closed and woody shrublands, savannas, and land used for grazing of livestock |
| Forest | Undisturbed or recovering habitat including evergreen, deciduous, needleleaf and broadleaf forests |
| Urban | Land converted to dense urban settlement |
| Barren | Bare or sparsely vegetated, including snow- and ice-covered areas and permanent wetlands. |

(b)

| Commodities | Associated land-use |
|---|---|
| Paddy rice | Cropland |
| Wheat | Cropland |
| Cereal grains | Cropland |
| Vegetables, fruits, nuts | Cropland |
| Oil seeds | Cropland |
| Sugar cane, sugar beet | Cropland |
| Plant-based fibres | Cropland |
| Other crops | Cropland |
| Cattle, sheep, goats, horses | Grassland & shrubland |
| Forestry | Forest |

Table 1: (a) Description of land use-classes used in the case studies; (b) GTAP commodities liked to land-use classes.

*Land use model:*

We adopted the land use modelling approach used by Kapitza *et al.* (2020). For Vietnam, spatial dynamics for land-use were simulated using the Dynamic Conversion of Land Use change and its Effects Model (Dyna-CLUE) (Verburg & Overmars 2009a) for Vietnam and using an R implementation (R package 'lulcc' (Moulds, Buytaert & Mijic 2015) of the Conversion of Land Use and its Effects at Small regional extents (CLUE-S) model (Verburg *et al.* 2002). Current land-use map for 2017 and 2019 (for Vietnam and Australia, respectively) were obtained from USGS (Broxton *et al.* 2014) and data were aggregated into five land-use classes (Table 1). To model land-use change, we first estimated total change in area per land-use class based on commodity demands from the GTAP model (Ha *et al.* 2017). We assumed that predicted increase in demand for an agricultural commodity in Vietnam under any scenario, would amount to a corresponding increase in area of in cropland, in a linear fashion, not accounting for increase in agricultural yields per unit area over time. Then, using a regression-based approach, the most likely places in the landscape for different land-use classes were identified based a number of environmental and anthropogenic variables (see Kapitza *et al.* 2020). Total area required per land-use class was then allocated across the study area based on the suitability for land-use classes within 1 km$^2$ grid cells and pre-determined conversion elasticities between land-use classes (Verburg & Overmars 2009a). Spatial interactions and competition between land use classes within grid cells were captured by conversion elasticities to reflect the costs of converting different land use type (Verburg *et al.* 2002; Verburg *et al.* 2004; Verburg & Overmars 2009) such that low values allow conversion between two types of land uses and high values constraints conversions (see Kapitza *et al.* 2020 for details on methods).

*Biodiversity model:*

Current and predicted land-use was used as an explanatory variable along with environmental variables within the biodiversity models ($R2 <= 0.7$). We used publicly available presence-only data on 747 and 639 bird species for Vietnam and Australia, respectively, from the Global Biodiversity Information Facility (www.gbif.org), and 1 km2 resolution climate and biophysical variables including topography, soil, elevation. Current and future bioclimatic variables were obtained from WorldClim Version 1.4 (Hijmans *et al.* 2005). Data on SSP trajectories used to define the SSP scenarios for Australia were obtained from the SSP public database hosted by the International Institute of Applied Systems Analysis (Riahi *et al.* 2017). Species distributions were modelled and predicted for 2070 using MaxEnt at 1 km2 resolution (dismo package version 1.1-4, Phillips, Anderson & Schapire 2006; Elith *et al.* 2011). The resulting maps indicate the suitability of currently available bird habitat in Vietnam and how this might change under future economic and environmental change, emphasising the risk of local extinctions due to lack of suitable habitat.

**Results**

*Case study I: Trade and bird biodiversity in Vietnam*

Results from the land use model indicate greater land use change under a business-as-usual scenario compared to the trade liberalization scenario (Figure 1). However, the directionality of change (in area of land use classes) in either scenario is the same. Under both the BAU and TPP scenarios, the land use model predicts substantial decrease in area of grass/shrubland land use class, whereas area of forest land use class under urbanization is expected to increase. The most significant change from grass/shrub to forest is expected in the northern region of Vietnam (Figure 2a). Corresponding changes in overall biodiversity (an area corrected richness indicator obtained from stacked distributions of all species for which a model could be fitted, 742 species) do not seem indicate any obvious distinctions in the patterns predicted under the two scenarios (Figure 2b). This is probably because individual species level differences can be lost when multiple species distributions are summarized into one composite indicator (Buckland *et al.* 2005).

Land use changes (Table 1b, Figure 2a) indicate potentially severe impacts for grassland compared to forest bird species for Vietnam. Vietnam is expected to witness transformation of rice paddies into secondary forests and plantations under a pacific trade agreement because (Ha *et al.* 2015) in some areas (Figure 1), thereby providing less habitat for grassland species in the future. For instance, the Brown Prinia *Prinia Polychroa*, which is primarily a ground and understory dwelling bird (Williams *et al.* 2014), is expected to lose more than 80% of its habitat under both scenarios, but the predicted decline under the BAU is higher compared to the TPP (Figure 3). Conversely, the Black-throated Laughingthrush (Garrulax chinensis) is expected to lose up to 50% of its habitat under the TPP scenario, which is expected to be much higher compared to the BAU scenario (Figure 3).

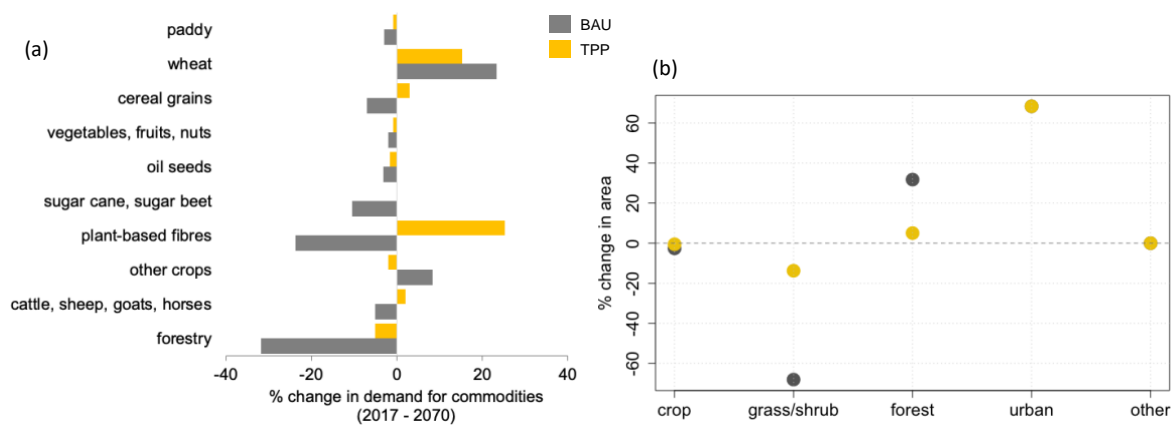

Figure 1: Predicted change in percentage of (a) commodity demand (Ha *et al.* 2017) and (b) area per land use class (commodities linked to land use classes as indicated in Table 1 and predicted using CLUEs model) in Vietnam over 2017 – 2070.

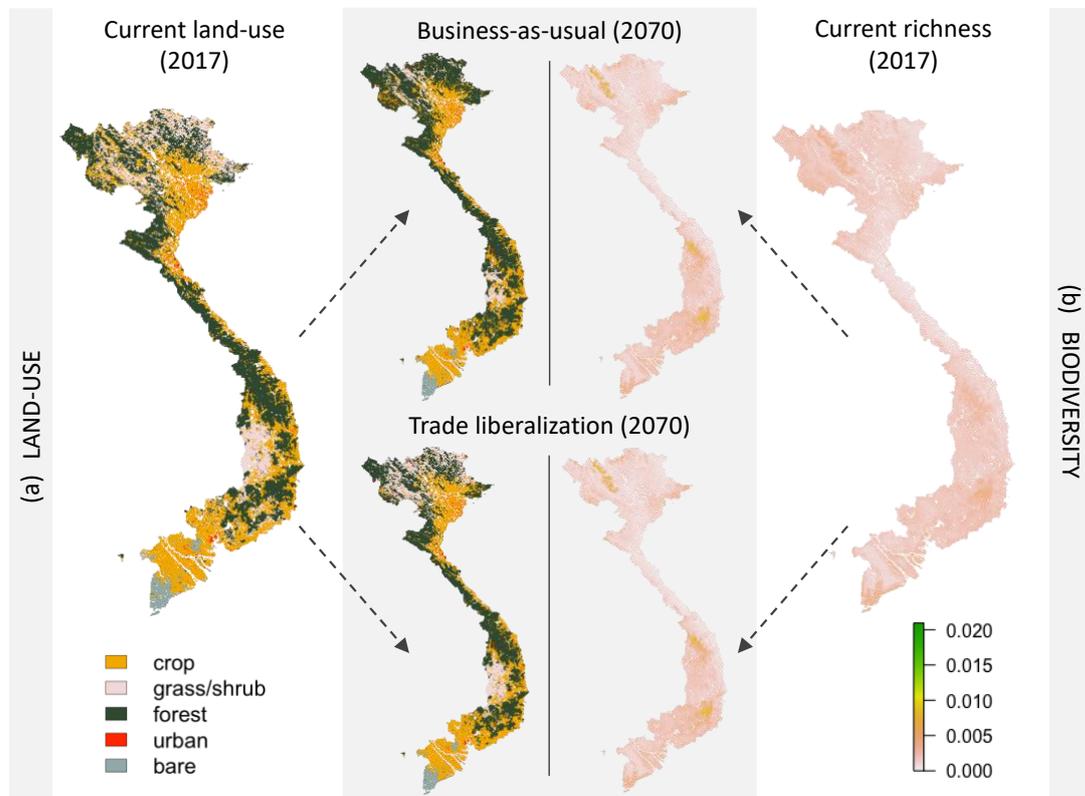

Figure 2: National scale analysis for Vietnam showing (a) land use and (b) biodiversity indicator (i.e., area weighted species richness) for 2017 and 2070 under the BAU and TPP scenarios. Biodiversity indicator is based on stacked species distributions for 742 bird species.

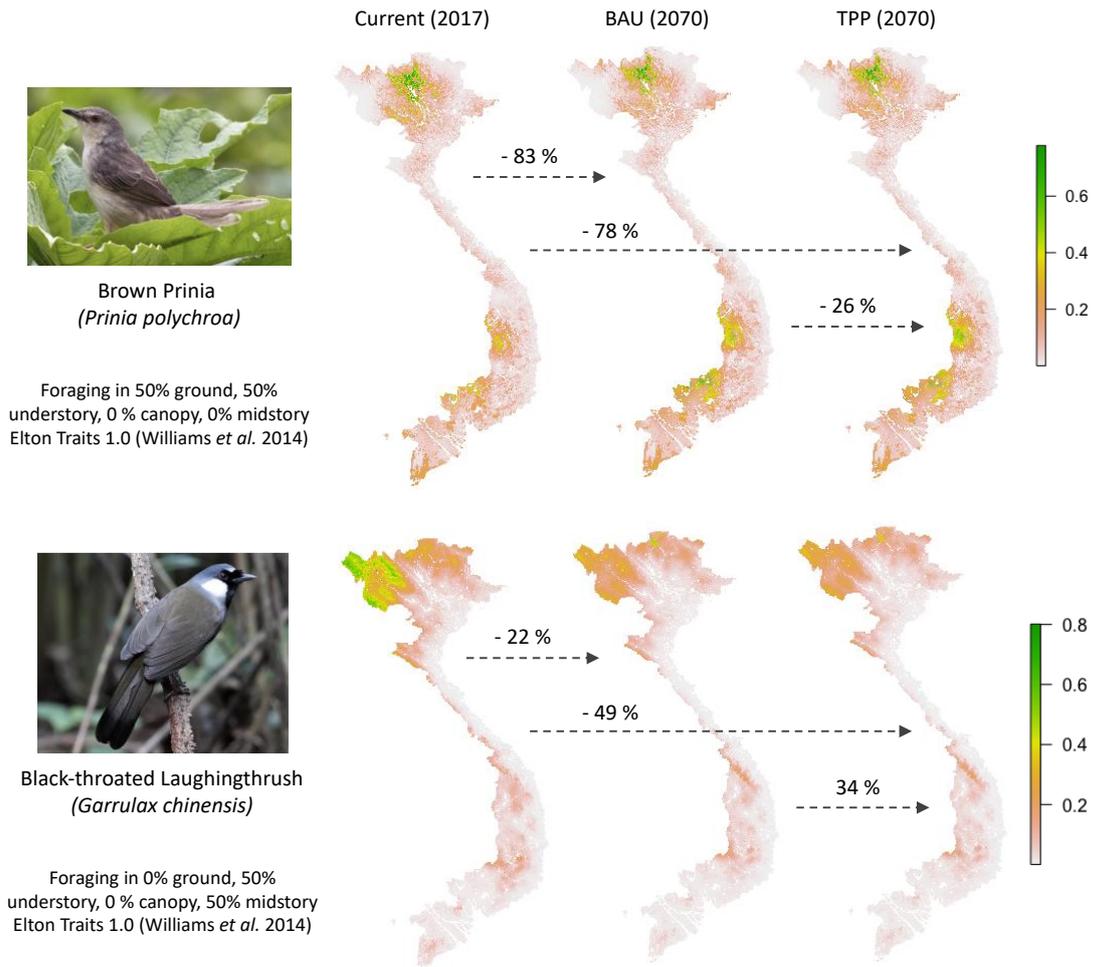

Figure 3: Habitat suitability maps for two bird species with distinct habitat under preferences (Williams *et al.* 2012) under the BAU and TPP scenarios. Numbers indicate the percentage decrease in habitat between scenarios indicated by the direction of the dotted arrows, e.g. the Brown Prinia habitat is predicted to decrease by 83% under BAU scenario from its current distribution, and the Black-throated Laughingthrush is expected to have 34 % more habitat area under the TPP as compared to the BAU scenario.

*Case study II: Biodiversity futures under coupled SSP-RCP scenarios in Australia*
In Australia, demand for wheat is expected to grow considerably while demand for oil seeds, livestock and forestry sectors is expected to drop under all three SSP-RCP by 2070 (Figure 4). Change in commodity demands drive change in the area required to produce these commodities (assumed linear relationship) within the land use model. However, predicted change in land use as per the *lulcc* model shows little difference amongst the scenarios (see example for SSP 3in Figure 5b) and only marginal differences from the current land use pattern (Figure 5c). A number of species show increase or decrease in habitat suitability but there are no distinguishable trends across the scenarios modelled (Figure 6a). Overall biodiversity trends can be represented by means of a biodiversity richness indicator (not shown here) from stacked species distributions of 639 bird species. Although land use appeared to be less important than the other variables considered in the model (Figure 6c), it was retained in 40% of the models. Figure 7 illustrates extent of declines in suitable habitats for example species which are predicted to lose over 90% of their current habitat (2019) under the coupled *regional rivalry* SSP3 – *extreme climate change* RCP 8.5 scenario.

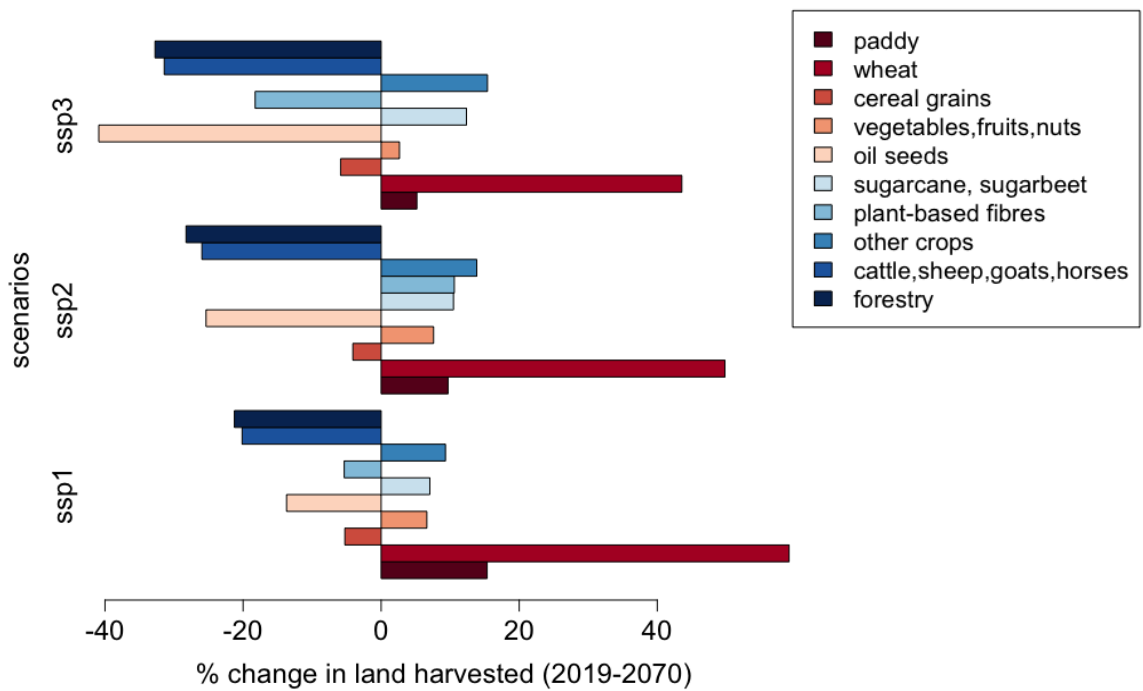

Figure 4: Change in land are (2019 – 2070) required to meet the projected commodity demands (see legend) under the three SSP scenarios.

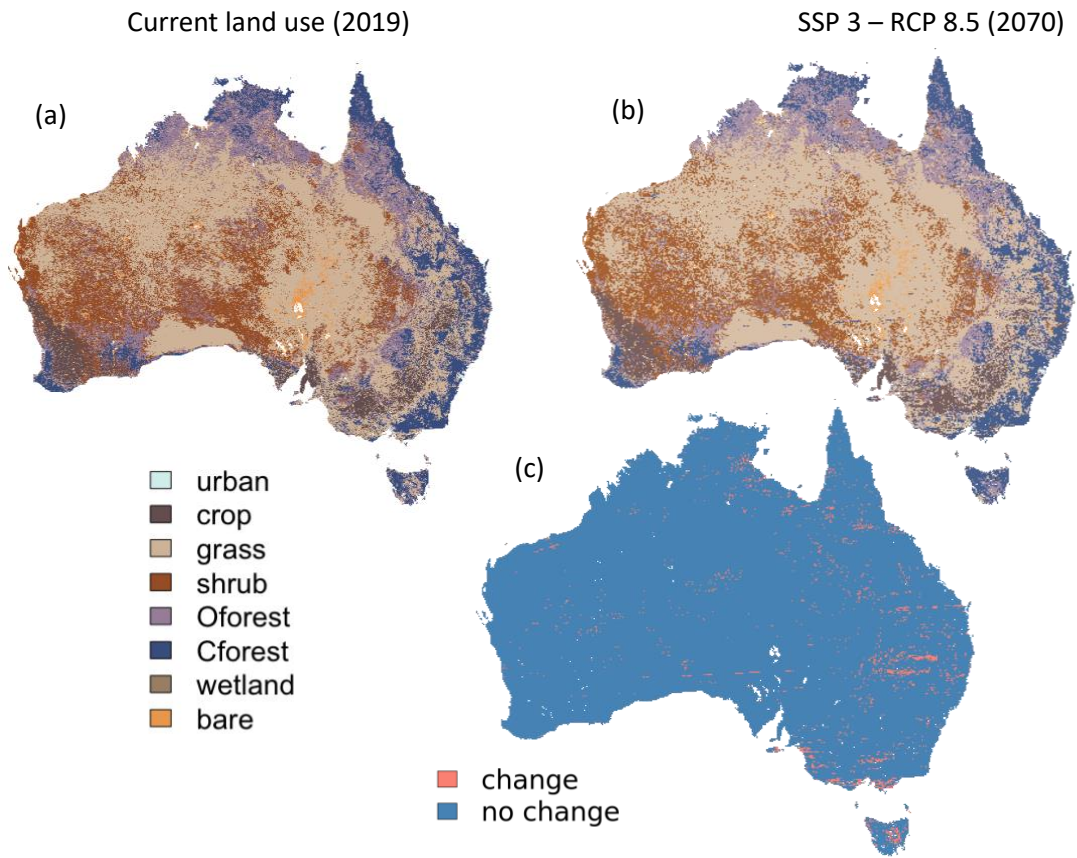

Figure 5: Land use predictions for Australia: (a) Current land use in Australia in 2019 (Brixton et al. 2014); (b) predicted land use under one of the three scenarios, i.e., the *'regional rivalry'* scenario with extreme climate change (SSP 3 – RCP 8.5), in 2070; and (c) change in land use from 2019 – 2070 under the SSP3 scenario.

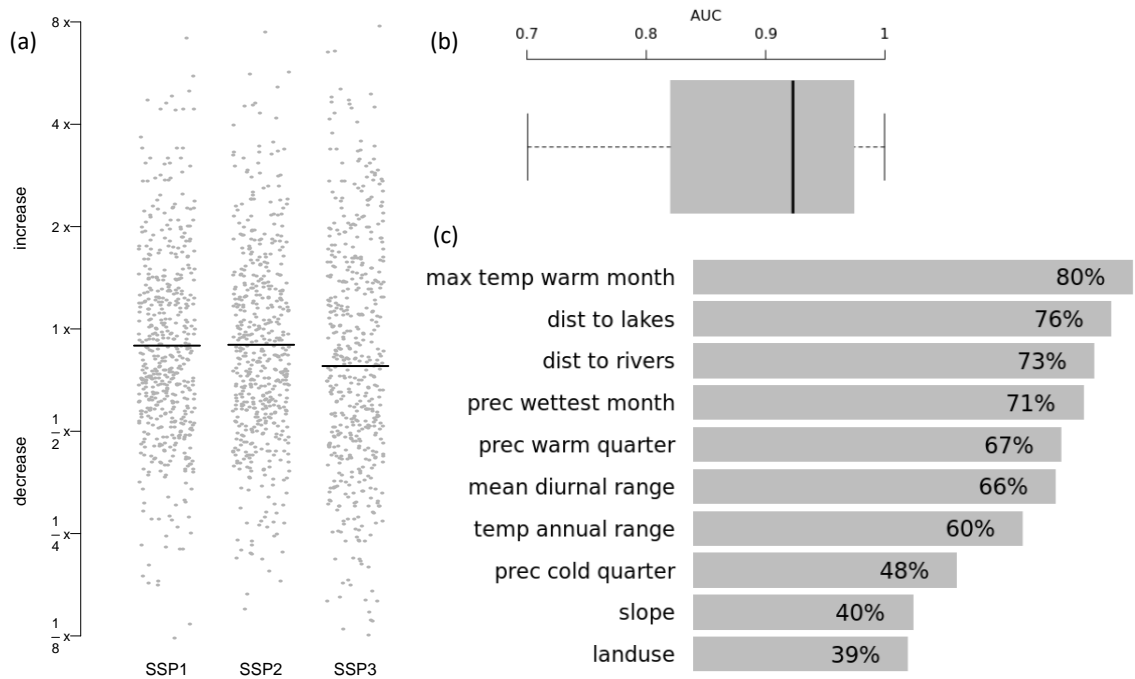

Figure 6: (a) Predicted change in species habitat for Australia (n = 639) from 2019 to 2070 under the three SSP – RCP scenarios: (1) low change SSP 1 – RCP 4.5, moderate change SSP 2 – RCP 6, and (3) high change SSP 3 – RCP 8.5. Each point corresponds to a species, black bars are the mean habitat change across all species. (b) AUC values for individual species models indicating predictive performance of the model retained (AUC > 0.7, n = 538). (c) Relative contributions of predictors in explaining differences between occupied and unoccupied locations (averaged across models). Numbers indicate the percentage of models in which a predictor variable was used.

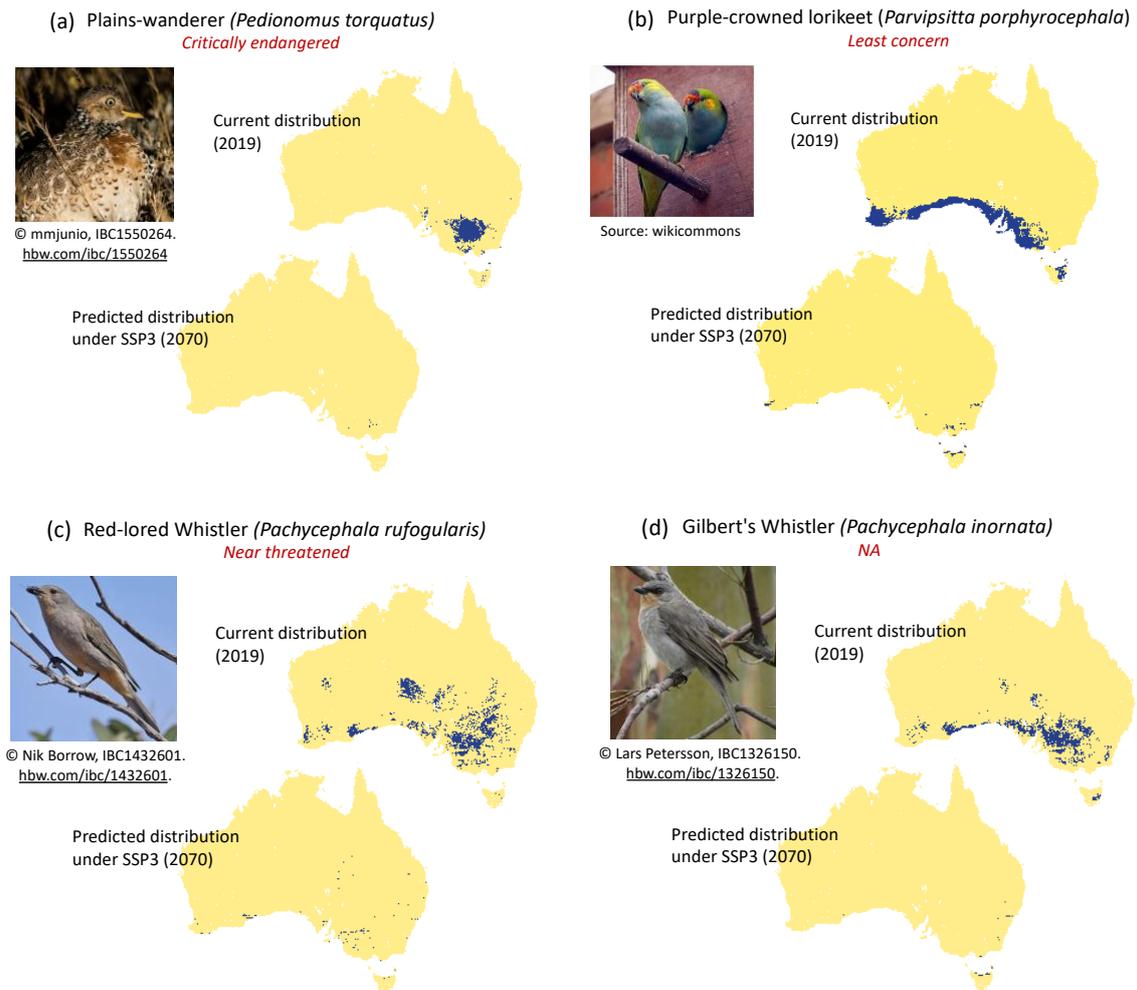

Figure 7: Habitat suitability for four bird species (a -d) in Australia showing > 90% decline in habitats from current (2019) to (mean) future predictions (2070) under the SSP 3 – RCP 8.5 scenarios representing future socio-economic and environmental conditions.

## *Opportunities of integrated socio-ecological analyses*

We provide the first integrated ecological-economic analysis pathway capable of supporting robust policy evaluation to identify critical risks and trade-offs for biodiversity and the economy under plausible global socio-economic and environmental futures at both national and global scales. In doing so, our study responds to repeated calls in the academic and policy literature for more integrated ecological-economic analyses to help support policies for environmentally sustainable development (Nash *et al.* 2020).

Shifting land use has been implicated in biodiversity loss at global scales (Foley *et al.* 2005; Newbold *et al.* 2019). Realistic exploration and prediction of future hotspots of land-use change and biodiversity impact are potentially extremely valuable for global organisations that seek to conserve biodiversity at a global scale. For example, The Nature Conservancy (TNC), the world's largest conservation land manager, seeks cutting-edge analyses that highlight places requiring special protection, or incentives for sustainable land use practices to compete with more damaging practices that may arise in response to fluctuating commodity demands in the global economy. The integrated framework proposed here allows the policy evaluation of incentives such as trade alliances as well as exploratory analysis the in the context of a shifting global land use profiles under alternate SSP scenarios for the economy and biodiversity. These analyses will help governments and large conservation organisations understand possible future biodiversity risks while there is time to act to minimise harm caused by emerging land use changes.

The environmental impacts of economic policies are seldom assessed beforehand, such that additional policies and mitigation measures often need to be put in place when unanticipated consequences arise. In this sense, environmental planning and policies are reactive rather than proactive (Cook *et al.* 2014). As FTAs continue to deal with minimizing "border" issues at the interface between two countries – tariffs and regulations that hinder imports, exports & investment – participating governments are becoming more concerned with "behind border" issues, particularly following policies like the North American Free Trade Agreement (NAFTA) in the 1990s. There is now a greater focus on including provisions for labour and human rights, environmental regulations, patent protection, and privacy laws before treaties can be ratified. These are long and arduous processes, with potentially far-reaching, long-term consequences; and to continue to draw them up without any quantitative measure of the environmental impacts of such treaties is not only politically inefficient, it is negligent. We have tailored an integrated modelling machinery to allow assessment of impacts of novel trading conditions, such as the emergence of new trade alliances between the US and the United Kingdom, or the outcomes of specific terms of the Trans-Pacific Partnership (TPP), and how those impacts interact with changing climates (also see Kapitza *et al.* 2020).

The analyses made possible by our proposed framework will provide insights into the environmental, social and economic implication of policy options for promoting broad scale vegetation restoration and carbon sequestration, as well providing tools to analyse any other policy options relating to other environmental challenges that have landscape scale land use implications. For example, our framework could be adapted to analyse the socio-economic and environmental benefits of changes to river regulation or environmental flows in the Murray-Darling Basin, to better understand the complete environmental-economic implications of agricultural incentives to reduce sediment and nutrient run-off to the Great barrier reef, or the economic and environmental costs and benefits of a significant national increase in hazard reduction burning. Similarly, global scale predictions of future land use and biodiversity conflict hotspots will be a key planning asset for large conservation organisations, as well as a valued contribution to global assessment processes such as the IPBES, providing input to Convention on Biological Diversity (CBD) targets (post-2020 Aichi targets) and UN Sustainable Development Goals 14 and 15 (Life Below Water and Life on Land).

### *Challenges & next steps*

Big data, sophisticated modelling approaches, combined with modern computing, make analyses possible that would've been considered unimaginable five years ago. Nonetheless, significant technical and conceptual challenges remain, particularly with respect to computation and the tight coupling of models to allow analyses of feedbacks and non-linearities that are so important when characterising the relationship between the economy and nature (Rosa *et al.* 2017).

*Giving life to SSPs*

Parameterisation of the SSPs for meaningful ecological-economic predictions requires greater detail within economic sectors (e.g. agriculture, forestry) and primary factors (necessary inputs for the production of intermediate and final goods, e.g. land, natural resources) in the CGE economics models. For instance, 'natural resources' such as fish stocks or mineral resources are considered as primary factors within a CGE, as capable of providing infinite supply; an assumption that can have grossly misleading consequences when CGEs are used to design environmental policy[19]. Furthermore, disaggregating forestry sector into managed and unmanaged, and agricultural land into cropland types, pastures and plantations is essential to model relevant land-use change capable of differentially impacting biodiversity. Close collaboration between the economics and computer science teams, will allow us to develop the high-dimensional analyses required to accommodate the larger number of sectors and additional constraints on model parameters (i.e. to limit supply for finite resources) within CGEs.

*Improving biodiversity metrics*

Modelling individual species provides some advantages over aggregated indices such as dissimilarity, richness or mean species abundance (MSA) (Ferrier *et al.* 2007) (Hui *et al.* 2013), particularly when national policies place an emphasis on individual (often threatened) species responses. However, other aggregate measures of biodiversity persistence could include variables such as the geometric mean extinction risk, the expected number of extinctions in the next 50 years (Nicholson & Possingham 2007), or the mean standardised expected minimum abundance (Wintle *et al.* 2011). Aggregate metrics from stacked species distributions, when individual species responses can be synergistic or antagonistic to drivers of change and declines, however, can cause biases in the biodiversity metric used to summarise overall biodiversity response and trends under the different scenarios. The choice of the metric/indicator and aggregation methods therefore needs further thought.

*Mechanisms of biodiversity impact*

The limited capacity of correlative models to act as a surrogate for extinction or persistence outcomes for species is widely acknowledged (Thuiller *et al.* 2004; Buckley & Roughgarden 2004; Harte *et al.* 2004). Mechanistic models of species population dynamics offer a promising alternative to understanding impacts of land-use change that overcome many of the shortcomings of correlative approaches (Akçakaya *et al.* 2004; Keith *et al.* 2008; Fordham *et al.* 2012). Mechanistic species persistence models expand the range of impacts on biodiversity that can be analysed because they address the mechanisms through which impacts occur, such as changes in species survival and fecundity. This provides a natural avenue for analysing impacts of land-use change and invasive species. However, these models require considerable data for parameterising and running models, which has constrained the number of species for which they can practically be implemented. The emergence of high quality, publicly available geo-spatial and biological databases (GEO BON, GBIF) open up new opportunities to rapidly produce species-level persistence analyses for thousands of species. The demand for such analysis in local, national and global sustainability assessments is now immediate and widespread (CBD3, IPBES4), while the technology and knowledge for delivering these outcomes are developing more slowly.

Finally, to address the issue of data deficient species, we will explore opportunities to draw predictive strength through multi-species modelling approaches, such as archetype models that cluster species based on their environmental response (Hui *et al.* 2013) and joint species distribution models (Wilkinson *et al.* 2019). The latter can also help account for biotic interactions between species that is currently not considered.

*Model integration (feedbacks and tight coupling)*
Socioeconomic and environmental interactions over distances can have profound implications for biodiversity and sustainability (Liu *et al.* 2013). Integrating spatial data and models in a clear framework enables reliable and reproducible analyses, while drawing on the best-available data from many real-world case studies, at otherwise infeasible scales. The current framework offers only a lose coupling of models such that outputs from 'upstream' models act as inputs for 'downstream' models. Full integration of the CGE economic model and the land use model in a way that that captures the multiple feedbacks between trade, climate, land constraints and endowments at the appropriate scale is challenging. There are many instances in which constraints on land endowment and elasticity of land use, along with the effects of climate and biosecurity events, limit the ability of the economy to produce commodities at all, much less at nominal production costs. This implies a dynamic equilibrium between production factors and land, a feature that is not currently accommodated in integrated assessment models, including our own. Hence, CGE models have not been able to adequately account for land use change and supply, and certainly not in a forward-looking, large-dimensional setting. As we progress this work, we will attempt to implement a dynamic feedback between our land allocation model and CGE model so that more realistic commodity production predictions can be made.

*Upscaling and downscaling (technical challenges of dimensionality)*

We aim towards building a fully integrated, tightly coupled ecological-economic model that will address the key challenges of dealing with feedback between modules (particularly economy and land endowments), and that will cope with the massive scale of computation and data for global analyses. A key challenge in this work, however, is the storage and access of large datasets (e.g. GBIF species occurrences, high resolution spatial environmental data) and seamless integration of outputs from the economic and land use models required to run species distribution models.

Coupling of land-use projections to biodiversity outcomes are currently rare and further innovation will be required to represent non-linear feedbacks between land-use decisions and commodity supply and demand. Addressing spatial autocorrelation for the land-use and biodiversity modelling sub-components at fine resolutions and/or at large extents is conceptually difficult (Guélat & Kéry 2018) and demands significantly greater computing resources. However, current computation size limits and long model runtimes make realistically complex analyses difficult to produce, refine, and validate. Finally, increasing the resolution of the analysis to include greater number of species and in land use projections by including land use classes meaningful for assessing biodiversity impacts (e.g. including multiple forest classes rather than just a single land class 'forest' (Hurtt *et al.* 2011; Rosa *et al.* 2020) will help align our work with inputs for earth system models and enable comparison of outputs across modelling approaches (Kim *et al.* 2018). We have ensured that our results are reproducible because the modelling platform are completely transparent, integrated (no manual porting of data between modules) and open access, and all data and modelling script are publicly available.

To address these challenges and accommodate the greater thematic, spatial and temporal resolution required to adequately test scenarios and policies of national and global relevance, we need to draw on cutting edge computational strategies and infrastructure. Innovative solutions for a) large data storage and efficient access; b) fast data analysis and modelling; and c) optimised and reusable pipelines are required as we move from local to planetary scales in an attempt to better address the grand challenge of building a sustainable future for nature and people. In keeping with this view, we have ensured that our results are reproducible because the modelling platform are completely transparent, integrated (no manual porting of data between modules) and open access, and all data and modelling script are publicly available.

**Conclusion**

Our work proves that the goal of predicting ecological outcomes of policy choices or global economic shocks can be achieved within an integrated ecological-economic modelling framework that incorporates both socio-economic drivers and environmental factors. These conceptual developments, combined with technical advances in computation indicate exciting new opportunities to provide high resolution economic, land-use and biodiversity predictions at spatial and temporal scales of immediate interest to policy makers. Needless to say, however, that further methodological advancements will be required to achieve the socio-ecological analysis at a spatial and commodity resolution relevant to land-use and biodiversity outcomes, and to achieve biodiversity analysis at the grain and extent required to understand the influence of trade, land-use and introduced species on persistence.